\documentclass[10pt,conference]{IEEEtran}
\usepackage{tabularx}
\IEEEoverridecommandlockouts
\usepackage[colorlinks,linkcolor=blue,anchorcolor=blue,citecolor=blue,urlcolor=purple]{hyperref}
\usepackage{cite}
\usepackage{amsmath,amssymb,amsfonts}
\usepackage{algorithmic}
\usepackage{graphicx}
\usepackage{textcomp}
\usepackage{booktabs}
\usepackage{xcolor}
\usepackage{pifont}
\usepackage{fontawesome}

\usepackage{amsmath,amssymb,amsfonts}
\usepackage{algorithmic}
\usepackage{algorithm}
\usepackage{array}
\usepackage{textcomp}
\usepackage{stfloats}
\usepackage{url}
\usepackage{verbatim}
\usepackage{graphicx}
\usepackage{cite}
\usepackage[colorlinks,linkcolor=blue,anchorcolor=blue,citecolor=blue,urlcolor=purple]{hyperref}
\usepackage{tcolorbox}
\usepackage{listings}
\usepackage{xcolor}
\usepackage[normalem]{ulem} 
\usepackage{graphicx}  
\usepackage{caption}
\usepackage{subcaption}
\usepackage{makecell}
\usepackage{color}
\usepackage{threeparttable}

\usepackage{float} 
\usepackage{pifont}
\usepackage{fontawesome}
\usepackage{multirow}
\usepackage{soul,framed}
\usepackage{cases}
\tcbuselibrary{most}
\usepackage{colortbl}
\usepackage{balance}

\def\BibTeX{{\rm B\kern-.05em{\sc i\kern-.025em b}\kern-.08em
    T\kern-.1667em\lower.7ex\hbox{E}\kern-.125emX}}
\begin{document}

\title{Towards Understanding Bugs in Distributed Training and Inference Frameworks for Large Language Models\\
} 

\author{\IEEEauthorblockN{Xiao Yu}
\IEEEauthorblockA{\textit{The State Key Laboratory of Blockchain and Data Security} \\
\textit{Zhejiang University}\\
Hangzhou, China \\
xiao.yu@zju.edu.cn}
\and
\IEEEauthorblockN{Haoxuan Chen}
\IEEEauthorblockA{\textit{School of Computer Science and Artificial Intelligence} \\
\textit{Wuhan University of Technology}\\
Wuhan, China \\
haoxuan.chen@whut.edu.cn}
\and
\IEEEauthorblockN{Feifei Niu}
\IEEEauthorblockA{\textit{School of Electrical Engineering and Computer Science} \\
\textit{University of Ottawa}\\
Ottawa, Canada \\
fniu2@uottawa.ca}
\and
\IEEEauthorblockN{Xing Hu} 
\IEEEauthorblockA{\textit{The State Key Laboratory of Blockchain and Data Security} \\
\textit{Zhejiang University}\\
Hangzhou, China \\
xinghu@zju.edu.cn}
\and
\IEEEauthorblockN{Jacky Wai Keung} 
\IEEEauthorblockA{\textit{Department of Computer Science} \\
\textit{City University of Hong Kong}\\
Hong Kong, China \\
jacky.keung@cityu.edu.hk} 
\and
\IEEEauthorblockN{Xin Xia}
\IEEEauthorblockA{\textit{The State Key Laboratory of Blockchain and Data Security} \\
\textit{Zhejiang University}\\
Hangzhou, China \\
xin.xia@acm.org}
}

\maketitle

\begin{abstract}

With the rapid advancement of large language models (LLMs), distributed training and inference frameworks such as DeepSpeed have become essential infrastructure for scaling both model training and inference across multiple GPUs or nodes. However, the growing complexity of these frameworks introduces non-trivial software bugs that can impair training performance, cause unexpected failures, and lead to significant resource waste. Understanding the characteristics of framework bugs is a foundational step toward quality assurance, enabling the design of more effective debugging and repair approaches. Therefore, our paper presents the first large-scale empirical study of 308 resolved bugs from three widely used distributed training and inference frameworks: DeepSpeed, Megatron-LM, and Colossal-AI. We examine bug symptoms, root causes, bug identification and fixing efforts, and common low-effort fixing strategies. Our findings show that the most frequent symptoms include crash, incorrect functionality, and build failure, while the bug root causes are primarily API misuse, incorrect implementation, configuration error, and missing preconditions. Additionally, the distributed nature of these frameworks introduces unique bug root causes, such as allocation strategy error and distributed communication error. Diagnosing and fixing complex bugs remains challenging due to factors like the disconnect between symptoms and root causes, high bug reproduction costs, and low-level or cross-component interactions. Interestingly, we observe that 48\% of bug fixes require minimal code changes ($\leq$ 10 LOC) and follow simple strategies such as conditional logic optimization, parameter handling enhancement, or version compatibility handling, indicating potential for automation. 
Based on these insights, we offer several implications for improving the reliability of both distributed training and inference frameworks and their dependent LLM projects, while also identifying opportunities to leverage LLM-based tools for automated debugging and repair.

\end{abstract}


\vspace{-5pt}
\section{Introduction}


In recent years, the field of artificial intelligence has witnessed exponential growth in model size, architectural complexity, and the volume of training data~\cite{zha2025data, yu2024makes}. This trend is particularly prominent in the development of large language models (LLMs), which often contain billions to hundreds of billions of parameters, necessitating distributed training and inference across multiple GPUs to meet computational and memory demands~\cite{naveed2023comprehensive, shanahan2024talking}. While general-purpose frameworks such as TensorFlow and PyTorch offer basic support for distributed training, they exhibit critical limitations when scaled to LLM-level workloads. These limitations include GPU memory saturation, sublinear scaling across nodes, and communication bottlenecks during gradient synchronization~\cite{gao2020estimating, jain2019scaling, dai2022reveal}. To address these challenges, specialized distributed training and inference frameworks, such as DeepSpeed~\cite{rasley2020deepspeed, deepspeed}, Megatron-LM~\cite{shoeybi2019megatron, megatronlm}, and Colossal-AI~\cite{li2023colossal, colossalai}, have emerged as essential infrastructure. These frameworks incorporate a comprehensive suite of optimization techniques, including advanced parallelization strategies (e.g., data, model, pipeline, and tensor parallelism) and memory optimization techniques (e.g., mixed-precision computation and ZeRO~\cite{rajbhandari2020zero}), to improve scalability and enhance efficiency in both training and inference while mitigating memory constraints.

As documented in Table \ref{tab:llm_frameworks}, the official GitHub repositories of these distributed training and inference frameworks explicitly list major LLMs trained using these frameworks, or specify the models they support for training/fine-tuning and inference. A representative example is the BLOOM~\cite{le2023bloom} model, which was trained using the Megatron-DeepSpeed framework. 
This composite solution integrates the Transformer implementation, tensor parallelism, and data loading from Megatron-LM, along with the ZeRO optimizer, model pipelining, and general distributed training components provided by DeepSpeed. The training process for BLOOM was highly resource-intensive, spanning 3.5 months and consuming 1,082,990 hours while utilizing 384 NVIDIA A100 80GB GPUs across 48 nodes~\cite{le2023bloom}. Such massive resource investments highlight the critical need for framework reliability, as latent bugs could degrade efficiency, induce process hangs, or trigger crashes, thus wasting costly computational resources. 

Despite their critical role, these frameworks are not immune to software bugs, as evidenced by real-world bug-related issues reported on GitHub. For instance, due to the absence of the \texttt{grad\_accum} attribute in DeepSpeed’s gradient accumulation process, \textit{OpenRLHF}~\cite{openrlhf} -- a prominent reinforcement learning from human feedback (RLHF) toolkit with 6,900 GitHub stars that relies on DeepSpeed as a core dependency---experienced unexpected training crashes during Llama model training. This bug~\cite{bunnydeepspeedbug} has drawn the attention of other developers. One developer described it as ``\textit{a serious bug},'' while another asked, ``\textit{Is there any workaround or solution can \textbf{quick} fix this issue}?'' Despite the severity and prevalence of bugs in distributed training and inference frameworks, no prior work has systematically characterized the nature of these bugs in such specialized systems. Understanding the characteristics of bugs in the frameworks is a fundamental step for quality assurance tasks, facilitating designing effective debugging and bug fixing approaches.
Therefore, we collect 308 resolved bugs from the GitHub repositories of DeepSpeed, Megatron-LM, and Colossal-AI (Table~\ref{tab:llm_frameworks}). We structure our study around the following four research questions (RQs):


\begin{table}[!t]
\centering
\caption{The statistics of the frameworks}
\renewcommand{\arraystretch}{1.2} 
\begin{tabular}{
>{\raggedright\arraybackslash}m{1.7cm}
>{\raggedright\arraybackslash}m{0.4cm}
>{\raggedright\arraybackslash}m{0.4cm}
>{\raggedright\arraybackslash}m{0.4cm}
>{\raggedright\arraybackslash}m{3.85cm}
}
\toprule
\textbf{Framework} & \textbf{Stars} & \textbf{Forks}  & \textbf{Bugs} & \multicolumn{1}{c}{\textbf{Supported LLMs}} \\ 
\midrule
DeepSpeed              & 38.1k          & 4.3k           & 168                                           & Megatron-Turing NLG-530b, BLOOM-176b, Jurassic-170b, ... \\ \hline

Megatron-LM        & 12.1k          & 2.7k           & 9                                             & Megatron-Turing NLG-530b,  BLOOM-176b, Retro-48b, ... \\\hline

ColossaI-AI     & 40.8k          & 4.5k           & 131                                            &  Open-Sora-11b, Stable Diffusion, Colossal-LLaMA-2-13b, ... \\ 
\bottomrule
\end{tabular} 
\label{tab:llm_frameworks}
\vspace{-17pt}
\end{table}

\noindent
\textbf{RQ1: What are the symptoms of bugs in the distributed training and inference frameworks?} Our analysis reveals that the most common symptoms are training crash, incorrect functionality, and build failure. In contrast, symptoms such as poor performance and CI/CD error occur less frequently, while system hang is the least commonly observed.

\noindent
\textbf{RQ2: What are the root causes underlying these bugs?} 
The most prevalent root causes include API misuse, configuration error, incorrect implementation, and precondition missing. Additionally, the distributed nature of these frameworks introduces several unique failure modes, including allocation strategy error, distributed communication error, distributed tensor error, and concurrency error. We observe that nearly all of these root causes can lead to the four most common symptoms: crash, incorrect functionality, build failure, and poor performance, highlighting the challenge of leveraging specific symptoms for bug root cause identification. 

\noindent
\textbf{RQ3: What are the key challenges that hinder effective root cause identification and bug fixing in distributed frameworks? }
6\% of the bugs remain open for over one month and have accumulated more than ten comments, indicating substantial debugging effort. Additionally, 20\% of the bugs require patch sizes exceeding 50 Lines of Code (LOC). Through in-depth analysis of these cases, we identify key challenges in root cause identification and bug resolution, including the disconnect between observed symptoms and their underlying causes, high computational overhead for bug reproduction, low-level architectural complexity, algorithm-specific logic errors, and cross-component interactions.

\noindent
\textbf{RQ4: What fixing strategies are commonly used for bugs resolved with minimal code changes?} 48\% of the bugs are resolved with small patches involving no more than ten LOC. These low-effort fixes typically follow common patterns such as conditional logic refinement, parameter handling enhancement, API replacement, configuration adjustment, and version compatibility handling. These strategies indicate strong potential for developing automated repair techniques in the future.

Based on these findings, we provide some practical implications for framework developers, users, and researchers. To mitigate framework bug risks, developers should strengthen API input validation, enhance version compatibility testing (e.g., PyTorch), and improve debugging support via mock modes (for lightweight reproduction) and context-rich logs (e.g., node IDs, tensor shapes). Framework users are recommended to adopt hybrid static–dynamic analysis tools to identify how bugs in the framework impact their downstream project components. The prevalence of small, recurring fixes suggests opportunities to combine template-based repair with LLMs. Additionally, domain-specific bugs (e.g., allocation strategy errors) and complex cross-component issues call for integrating retrieval-augmented generation (RAG) with reasoning-oriented LLMs (e.g., GPT-o1~\cite{OpenAIO1}, DeepSeek-R1~\cite{guo2025deepseek}) and program dependency analysis for systematic resolution.

Our paper makes the following key contributions:

\noindent
(1) We present the first large-scale empirical study characterizing 308 bugs across three widely used distributed training and inference frameworks. Furthermore, we make our annotated dataset publicly available to facilitate future research~\cite{bugissuedataset}.
    
\noindent
(2) Our analysis uncovers major challenges in diagnosing and resolving complex bugs, along with common low-effort fixing strategies that are potentially automatable.

\noindent
(3) We derive implications for enhancing the reliability of distributed training and inference frameworks and their downstream LLM projects, while also identifying opportunities to leverage tools for automated debugging and repair.


\section{Study Design}
\begin{figure}[t]
    \centering
    \includegraphics[width=\columnwidth]{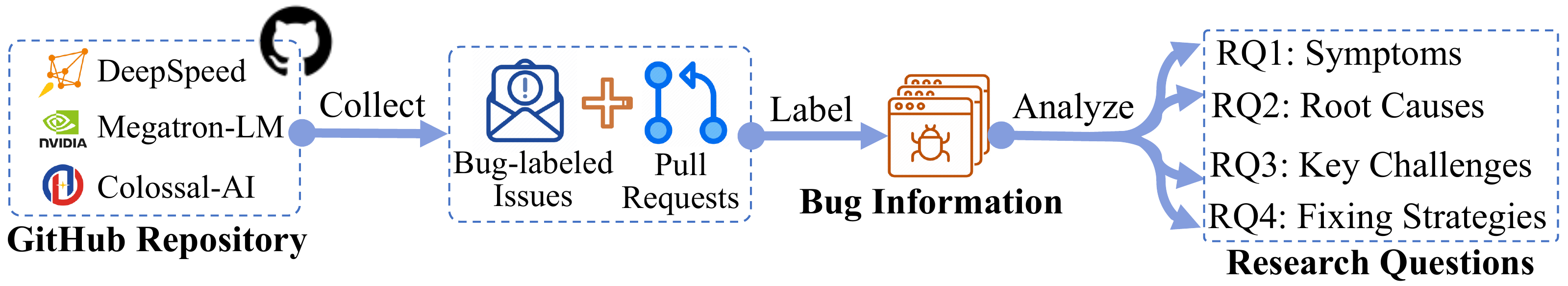} 
    \caption{The overview of our study design.}
    \label{fig:study_overview_design}
       \vspace{-17pt}
\end{figure}

\label{sec: study design}
The overview of our study design is shown in Figure \ref{fig:study_overview_design}. 
\subsection{Framework Selection}

To identify representative distributed training and inference frameworks, we define three selection criteria. First, following prior research on analyzing deep learning framework bugs~\cite{chen2023toward,yang2025towards}, we prioritize actively maintained distributed training and inference frameworks with substantial community engagement, measured by GitHub stars and forks. 
Second, we require official documentation that explicitly lists major LLMs trained using these frameworks or the specific models they declare support for training or fine-tuning. Finally, we include only frameworks with a history of closed issues that are labeled as ``bug'' and linked to resolved pull requests (PRs). 
Several distributed training and inference frameworks fail to meet our selection criteria for the following reasons: EasyParallelLibrary~\cite{jia2022whale,easyparallellibrary} and Alpa~\cite{alpalibrary} have not been updated for over two years; MindSpeed-LLM~\cite{mindspeed-llm} has very few stars and forks; and Meta Lingua~\cite{metalinguaarticle, lingualibrary} lacks closed issues labeled as ``bug''. Ultimately, we select three widely used frameworks: DeepSpeed (Microsoft)~\cite{rasley2020deepspeed, deepspeed}, Megatron-LM (NVIDIA)~\cite{shoeybi2019megatron, megatronlm}, and Colossal-AI (HPC-AI Tech)~\cite{li2023colossal, colossalai}. These frameworks adopt a comprehensive suite of optimization techniques to improve scalability, efficiency, and memory usage during distributed training or inference. For example, DeepSpeed exemplifies such advancements through a core architecture integrating several sophisticated strategies. 

(1) \textbf{3D Parallelism}:
DeepSpeed introduces 3D parallelism, a hybrid strategy that combines data parallelism, model parallelism, and pipeline parallelism. Data parallelism replicates the model across multiple devices and trains on different batches of data. Model parallelism splits the model across devices to accommodate extremely large architectures. Pipeline parallelism divides the model into sequential stages and processes micro-batches in a pipelined fashion across devices. By integrating these three forms of parallelism, DeepSpeed achieves high scalability across thousands of GPUs while maintaining strong throughput and minimizing communication overhead.

(2) \textbf{Memory Optimization Techniques}: 
In traditional settings, each GPU maintains a complete copy of the model states, including parameters, gradients, and optimizer states, which significantly limits scalability. ZeRO~\cite{rajbhandari2020zero} mitigates this issue by partitioning these components across data-parallel processes, consisting of three stages: Stage 1 partitions optimizer states; Stage 2 adds gradient partitioning; and Stage 3 partitions the model parameters. Mixed-precision training (e.g., using fp16/fp8 formats) lowers memory overhead and accelerates computation. Gradient accumulation simulates larger batch sizes without increasing memory usage, while CPU offloading transfers parts of the optimizer state or gradient computations to CPU memory, alleviating GPU memory pressure. Activation checkpointing reduces memory consumption by saving only a subset of intermediate activations during the forward pass and recomputing them during backpropagation.

\subsection{Data Collection}

To systematically investigate the characteristics of bugs in distributed training and inference frameworks, consistent with prior analyses of bugs in machine learning and deep learning frameworks~\cite{li2023understanding, ho2023empirical, tambon2024silent, long2022reporting, yang2025towards}, we analyze closed issues labeled as ``bug'' that are explicitly linked to merged PRs in each framework’s repository. A closed issue signifies formal acknowledgment and resolution of the bug by the development community. We adopt this issue-based approach rather than directly analyzing bug-fixing PRs for four key reasons:
(1) PRs often contain feature enhancements, performance optimizations, or other non-defect changes, making it difficult to identify bug fixes without explicit linkage to issues. 
(2) Many bug-fixing PRs (e.g.,~\cite{deepspeedpr3609,colossalaipr1299}) lack sufficient textual information, making it difficult to infer the symptoms and root causes of the bugs. 
(3) In contrast, issue descriptions typically document the failures encountered by developers, such as training crashes, offering valuable context on the real-world symptoms of bugs during LLM training. 
(4) The number of comments on a bug-labeled issue, along with its timestamps (i.e., creation and closure dates), enables us to evaluate both the difficulty of root cause identification and the bug’s open duration. The ``Bugs'' column in Table~\ref{tab:llm_frameworks} reports statistical details on the number of closed bug-labeled issues across the three distributed training and inference frameworks, with each issue linked to a corresponding merged PR. We also confirm that all bugs in our dataset are unique, ensuring that each entry represents a distinct bug instance analyzed in this study.

\subsection{Manual Labeling of Bug Symptoms and Root Causes}
 \label{sec: Manual Labeling of Bug Root Causes and Symptoms}

We systematically label each bug’s symptom, root cause, and occurrence phase (training or inference) through a two-stage process. First, we randomly sample 40\% of the collected bug-labeled issues and their PRs for pilot labeling to develop initial taxonomies.
Specifically, following the open-coding scheme~\cite{seaman1999qualitative}, two authors adapt general symptom/root cause taxonomies from prior deep learning framework bug studies~\cite{chen2023toward, ho2023empirical, li2023understanding}. They refine these taxonomies by adding framework-specific categories (e.g., distributed communication error) and removing irrelevant ones. An independent third researcher with expertise in distributed training and inference frameworks arbitrates disagreements via consensus discussions, finalizing the taxonomies before formal labeling begins. 
For formal labeling of the remaining 60\% of the dataset, the two authors independently label in four rounds (each 15\% of the total dataset), measuring inter-rater agreement via Cohen’s Kappa~\cite{cohen1960coefficient} after each round. The first round yields a Kappa score of 0.5, prompting thorough discussions with the arbitrator to resolve inconsistencies and clarify labeling criteria. The second round achieves a Kappa score greater than 0.8, while the final two rounds exceed 0.9, with the arbitrator continuously mediating all round-specific disagreements to ensure consistent labeling across all data.

\subsection{Challenges in Bug Root Cause Identification and Fixing, and Common Fixing Strategies}

We quantify bug resolution effort using three indicators widely adopted in prior work~\cite{zaman2012qualitative, liu2014characterizing, yang2025towards}: 
(1) \textit{Bug Open Duration}, defined as the interval between the opening and closing of a bug-labeled issue;
(2) \textit{Number of Comments}, representing the volume of discussions among developers and users during the bug-labeled issue lifecycle;
(3) \textit{Patch Size}, measured by the number of LOC modified to fix the bug. To ensure meaningful measurement, we exclude changes to non-production files such as tests, benchmarks, and documentation, and remove blank lines from the LOC count.

Following previous work~\cite{wang2021exploratory,yang2025towards}, we analyze bug-labeled issues that remain open for at least one month and receive more than 10 comments to investigate the challenges in root cause identification. To explore high-effort fixes, we examine bugs whose patch sizes fall in the top 20\% of our dataset, identifying common difficulties in resolving complex bugs. Additionally, we analyze bug fixes with patch sizes no greater than 10 LOC and identify several common low-effort fixing strategies that present opportunities for automation.
The manual labeling of challenges in bug root cause identification and bug resolution, as well as the classification of common fixing strategies, follows a procedure analogous to the symptom and root cause labeling process described in Section \ref{sec: Manual Labeling of Bug Root Causes and Symptoms}.

\section{Results}
\label{sec: results}
\subsection{RQ1: Symptoms Analysis}



\begin{figure}[h]
   \vspace{-13pt}
    \centering
    \includegraphics[width=\columnwidth]{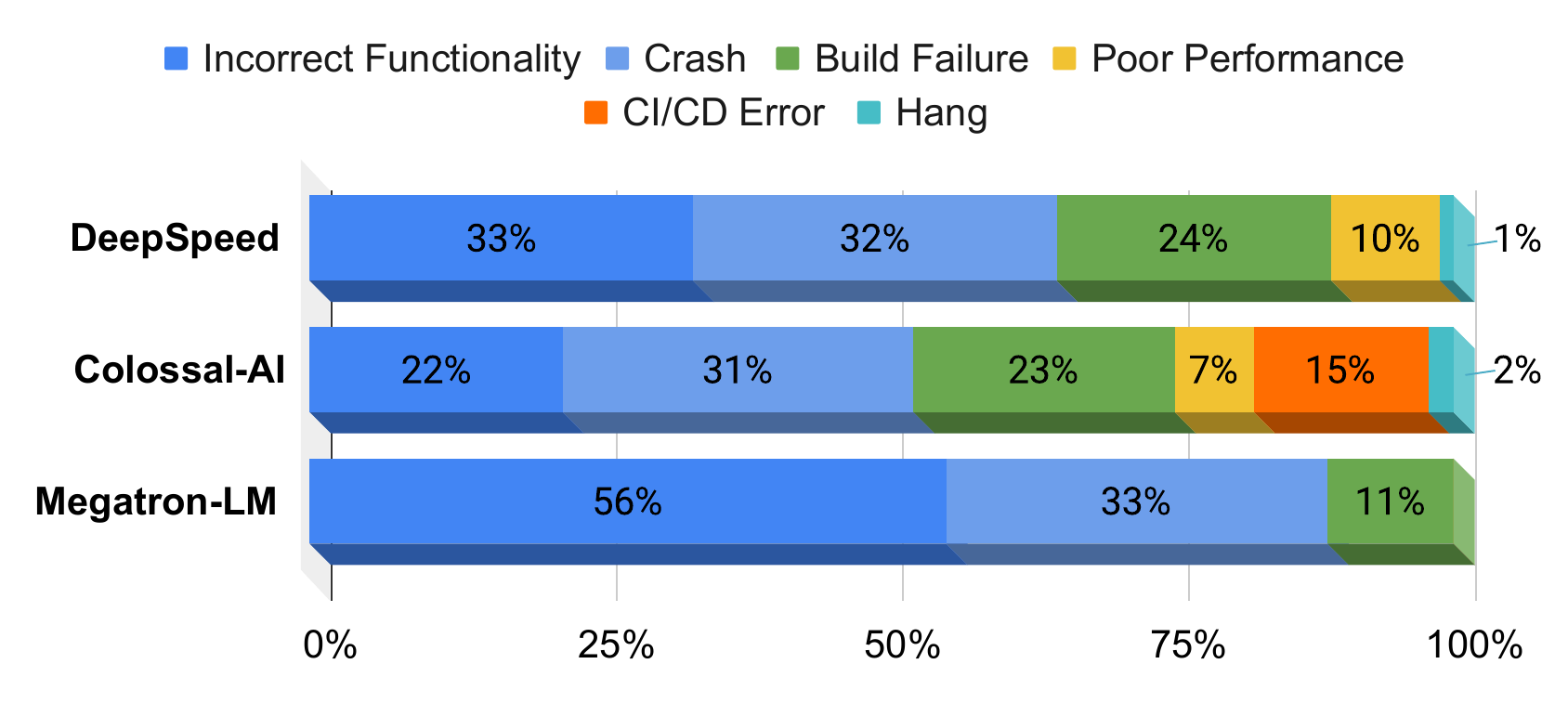} 
    \caption{The bug distribution by symptoms.}
    \label{fig:bug_symptoms}
       \vspace{-15pt}
\end{figure}

\begin{figure*}[h]
    \centering
    \includegraphics[width=\textwidth]{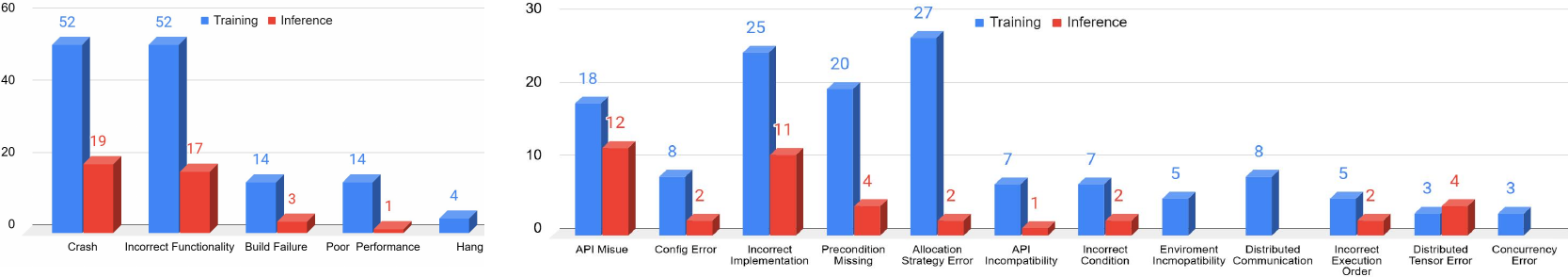}
    \caption{The number of bugs occurring during the model training and inference phases across the three frameworks.}
\label{fig:train_inference_symptom_root_causes_distribution}
\vspace{-10pt}
\end{figure*}

Figure \ref{fig:bug_symptoms} illustrates the distribution of bugs by the identified symptoms across the three distributed training and inference frameworks. Notably, crash, incorrect functionality, and build failure emerge as the three most common symptoms across these frameworks. 
Figure \ref{fig:train_inference_symptom_root_causes_distribution} shows the number of bugs of the three framework occurring during the model training or inference phases. Since some bug-labeled issues lack sufficient information to determine whether the bug occurred during training or inference, these bugs are excluded from the statistics presented in Figure \ref{fig:train_inference_symptom_root_causes_distribution}. The number of bugs related to crashes, incorrect functionality, build failures, and poor performance during the training phase is more than two times higher than that during the inference phase. Furthermore, hang bugs are found exclusively during the model training phase.

\noindent\faBug\ \textbf{Crash.} This type of symptom means that the program can be compiled and launched successfully, but it terminates unexpectedly during execution due to underlying bugs. For example, in DeepSpeed Issue \textit{\#3543}~\cite{deepspeed3543}, the mismatch of tensors was not considered, resulting in an error during the LoRA process and an abnormal termination of RLHF. In Megatron-LM Issue \textit{\#437}~\cite{megatronlm437}, a bug in the \texttt{view} operation during \textit{Grouped Query Attention} computation led to a runtime crash due to incompatible tensor shape/stride mismatches when pipeline parallelism is enabled. In Colossal-AI Issue \#\textit{5187}~\cite{colossalai5187}, a custom shard policy caused a crash by triggering a \texttt{TypeError} due to layer distribution inconsistencies.

\noindent\faBug\ \textbf{Incorrect Functionality.} This type of symptom refers to scenarios where the program runs normally but produces incorrect output. For example, in DeepSpeed Issue \textit{\#1950}~\cite{deepspeed1950}, 
the program executed successfully; however, repeated runs with identical inputs and a fixed seed yielded non-deterministic inference results. 
In Colossal-AI Issue \textit{\#3746}~\cite{colossalai3746}, the prompt dataloader failed to properly iterate through batches, causing the program to repeatedly load identical prompts.




\noindent\faBug\ \textbf{Build Failure.} This category of symptoms refers to situations where the program fails to complete the compilation process due to coding errors or environmental issues during the compilation phase. For example, in DeepSpeed Issue \textit{\#3364}~\cite{deepspeed3364}, DeepSpeed did not support Apple M2 chips at the time, making it impossible to install and compile on such devices. In Colossal-AI Issue \textit{\#3620}~\cite{colossalai3620}, the script \texttt{train\_prompts.sh} failed to execute with a single GPU.

\noindent\faBug\ \textbf{Poor Performance.} This category of symptoms refers to situations where the program runs correctly and produces expected results, but execution time is significantly longer than normal, or it consumes excessive memory. For example, in Colossal-AI Issue \textit{\#5937}~\cite{colossalai5937}, the program saved checkpoints for each epoch, leading to wasted disk space. In DeepSpeed Issue \textit{\#1635}~\cite{deepspeed1635}, optimization flaws caused the execution time of the Adam optimizer on AVX512 instruction sets to increase over four times, severely impacting training efficiency.


\noindent\faBug\ \textbf{CI/CD Error.} This type of symptom indicates bugs in the library's release process that prevent users from obtaining the latest version in a timely manner. For example, in Colossal-AI Issue \textit{\#3848}~\cite{colossalai3848}, the Docker image push stage within the CI/CD pipeline was inadvertently skipped due to misaligned workflow triggers and conditional checks, preventing users from accessing the program’s latest Docker image.
Notably, CI/CD errors are unique to Colossal-AI because DeepSpeed treats CI/CD errors as non-bug issues under the dedicated label ``CI Failed'' with only nine closed issues that do not directly correlate with PRs, while Megatron-LM has relatively few issues with no recorded CI/CD errors to date. 


\noindent\faBug\ \textbf{Hang.} This symptom occurs when the program stalls at a certain execution stage due to specific reasons, without producing any output or prompt messages, making it difficult to determine whether the program is still running correctly. For example, in DeepSpeed Issue \#\textit{1995}~\cite{deepspeed1995}, the program failed to properly handle termination signals, causing it to become a zombie process. In Colossal-AI Issue \#\textit{4164}~\cite{colossalai4164}, configuration parameters interfered with the coordination mechanism between distributed training nodes, leaving the program indefinitely stuck in the multi-node initialization phase.

\begin{center}
\vspace{-15pt}
    \resizebox{\linewidth}{!}{
\begin{tabular}{l!{\vrule width 1pt}p{0.9\columnwidth}}
    \makecell{{\LARGE \faLightbulbO}}  &\textbf{Finding 1.} Crash, incorrect functionality, and build failure are the three most common symptoms in these three distributed training and inference frameworks, while hang occurs the least frequently. 
\end{tabular}}
\vspace{-8pt}
\end{center}

\subsection{RQ2: Root Cause Analysis}

\begin{figure*}[h]
    \centering
    \includegraphics[width=0.95\textwidth]{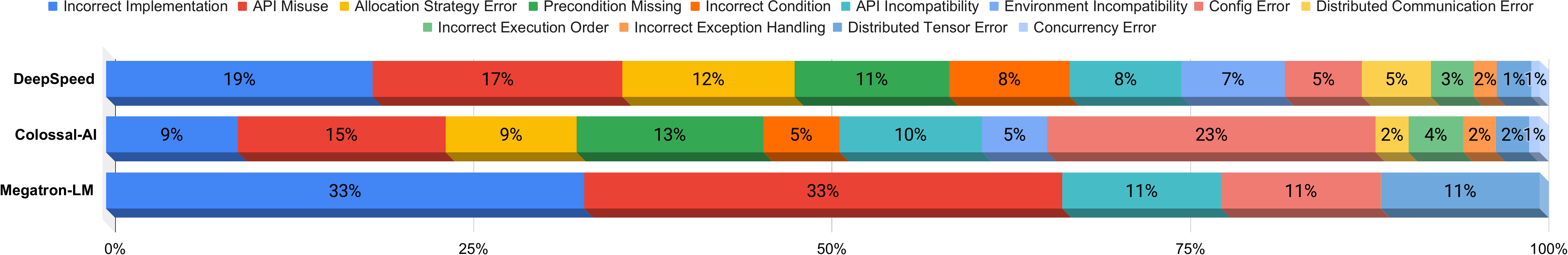}
    \caption{The bug distribution by root causes (Percentages are rounded, so totals for each project may not add up to 100\%).}
    \label{fig:bug_code_reason_static}
    \vspace{-20pt}
\end{figure*}

\begin{figure}[t]
    \centering
    \includegraphics[width=\columnwidth, height=0.5\textheight]{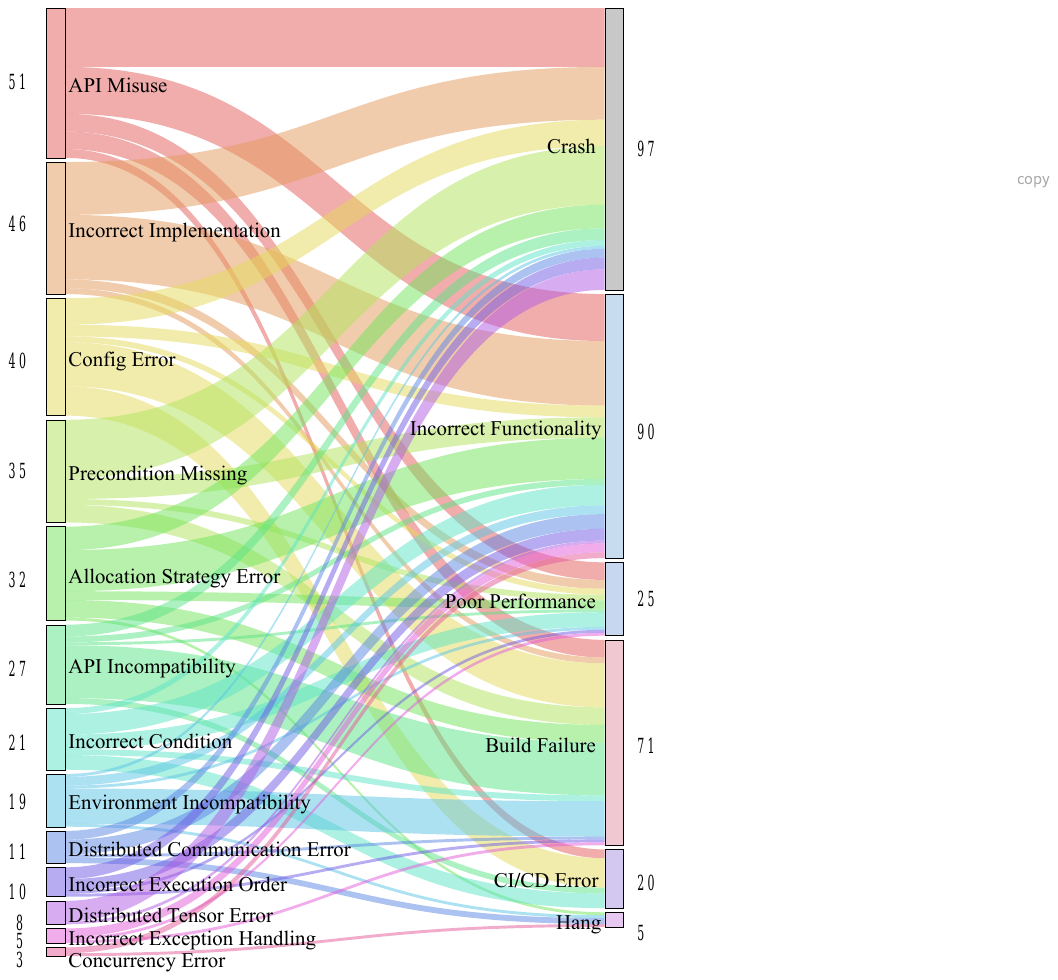} 
    \caption{The relationship between root causes and symptoms.}
    \label{fig:root_cause_symptom_relationship}
    \vspace{-20pt}
\end{figure}

Figure \ref{fig:bug_code_reason_static} illustrates the distribution of bugs categorized by their root causes across the three frameworks. Notably, API misuse, configuration error, incorrect implementation, and precondition missing emerge as the four most prevalent root causes. As shown in Figure \ref{fig:train_inference_symptom_root_causes_distribution}, 
while API misuse, incorrect execution order, and distributed tensor error result in a similar number of bugs during both the training and inference phases, other root causes lead to a noticeably higher number of bugs during the training phase. 
Figure \ref{fig:root_cause_symptom_relationship} demonstrates the correlations between root causes and symptoms, revealing how specific root causes contribute to diverse symptom manifestations while certain symptoms are concentrated around root causes. Because these frameworks are designed for distributed training and inference, which features complex resource allocation, node communication, and tensor parallelism, four distinct root causes arise: allocation strategy error, distributed communication error, distributed tensor error, and concurrency error. Due to space constraints, we present only source code examples for these four root cause categories to facilitate a more detailed explanation of these root causes.

\noindent\faBug\ \textbf{API Misuse: }This error often occurs due to developers fail to fully consider the parameter design, invocation methods, or underlying implementation mechanisms of APIs during design or usage. For example, in DeepSpeed Issue \textit{\#2244}~\cite{deepspeed2244}, oversight of permission constraints in the \textit{os.makedirs} function resulted in unwritable directories. In Megatron-LM Issue \textit{\#624}~\cite{megatronlm624}, essential parameters were omitted when calling the \textit{load} function, preventing test suites from executing properly.

\noindent\faBug\ \textbf{Config Error:} This error arises from configuration flaws in fixed parameters within library code, configuration files, or command scripts, causing anomalies during program installation and execution. For example, in DeepSpeed Issue \textit{\#11}~\cite{deepspeed11}, the multi-node installation script defaulted to a shared file system and lacked parameters for customizing the hostfile path, rendering it impossible to distribute installation files across nodes in non-shared file system environments and causing installation failures. In Colossal-AI Issue \textit{\#4027}~\cite{colossalai4027}, setting the inter-node communication bucket size to 12T in code erroneously degraded runtime efficiency.

\noindent\faBug\ \textbf{Incorrect Implementation: }This type of error arises from developers' incorrect application or implementation of technical solutions in function/algorithm realization. For example, in DeepSpeed Issue \textit{\#1950}~\cite{deepspeed1950}, logical inconsistencies in positional encoding and fused computation during inference kernel implementation caused non-deterministic inference results. In Colossal-AI Issue \textit{\#3063}~\cite{colossalai3063}, the \textit{OPTActor} module was initialized without properly loading pretrained weights, leading to a mismatch between the custom model's architecture and the pre-trained model's state dictionary.

\noindent\faBug\ \textbf{Precondition Missing:} This error occurs when critical preconditions are not checked before executing subsequent logic, leading to exceptions or system crashes under specific scenarios. For example, in DeepSpeed Issue \textit{\#4095}~\cite{deepspeed4095}, developers failed to detect the case where the input parameters are empty when creating tensors, causing the program to crash directly when the code processes the creation of empty tensors.
In Colossal-AI Issue \textit{\#3620}~\cite{colossalai3620}, the program failed to handle the single-GPU precondition (i.e., \textit{dist.get\_world\_size()} == 1), leaving data sampler variables undefined and causing initialization failures during single-node deployment.

\begin{figure}[h]
    \centering
    \includegraphics[width=\columnwidth]{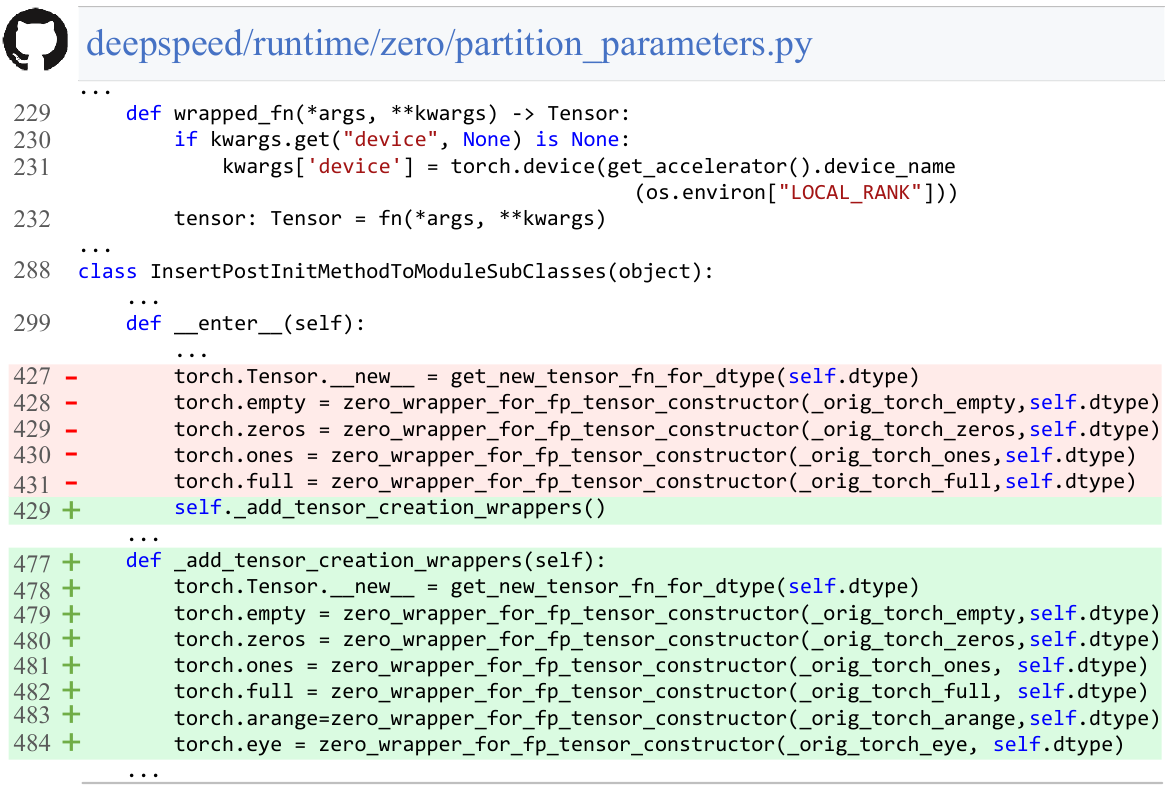} 
    \caption{An example of the allocation strategy error.}
    \label{fig:bug_allocation_strategy_error}
    \vspace{-20pt}
\end{figure}
\noindent\faBug\ \textbf{Allocation Strategy Error:} This type of error primarily results from logical flaws in implementing resource allocation strategies for parameters, optimizer states, or memory. For example, in DeepSpeed Issue \textit{\#3678}~\cite{deepspeed3678}, as shown in Figure \ref{fig:bug_allocation_strategy_error}, in the wrapper function \texttt{wrapped\_fn} for tensor creation, the current device (via \texttt{LOCAL\_RANK}) was dynamically retrieved and the device parameter was injected to ensure tensors are created on the target GPU (lines 229–232). The original code only wrapped functions such as \texttt{empty}, \texttt{zeros}, \texttt{ones}, and \texttt{full} (lines 427-431), but did not handle \texttt{arange} (arithmetic tensor) and \texttt{eye} (identity matrix). When the two functions were called without explicitly specifying the device, tensors were created on the CPU by default, leading to device mismatch. Therefore, the fixed code added wrappers for \texttt{arange} and \texttt{eye} (lines 483–484). 
In ColossalAI Issue \#\textit{5187}~\cite{colossalai5187}, inconsistent method invocations caused custom layer distribution logic to fail during pipeline forward passes, leading to mismatched layer indices and runtime errors.

\noindent\faBug\ \textbf{API Incompatibility:} This type of incompatibility can be divided into internal API incompatibilities and third-party package API incompatibilities. Internal incompatibilities arise when API changes within a library are not updated across all usage points. For instance, in DeepSpeed Issue \textit{\#7014}~\cite{deepspeed7014}, a class rename and method signature changes were not fully synchronized, causing runtime errors due to mismatched arguments. Third-party incompatibilities mainly stem from PyTorch version updates. Examples include DeepSpeed Issue \textit{\#5603}~\cite{deepspeed5603}, Issue \textit{\#5534}~\cite{deepspeed5534}, Issue \textit{\#4853}~\cite{deepspeed4853} and Colossal-AI Issue \textit{\#4829}~\cite{colossalai4829}, Issue \textit{\#2938}~\cite{colossalai2938}, all due to delayed adaptation to PyTorch version changes.

\noindent\faBug\ \textbf{Incorrect Condition:} This error type results from flawed or incomplete conditional checks in branch statements, causing unexpected program termination under specific conditions. In DeepSpeed Issue \textit{\#3751}~\cite{deepspeed3751}, a wrong conditional check during weight transfer logic prevented operation completion. In Colossal-AI Issue \textit{\#3237}~\cite{colossalai3237}, the CI/CD workflow used the undefined \textit{steps.commit.outputs.status} in conditions, causing the workflow to never trigger as the condition always failed. 

\looseness=-1
\noindent\faBug\ \textbf{Environment Incompatibility:} This error arises when developers overlook compatibility between the library and hardware architecture. For example, in DeepSpeed Issue \textit{\#3364}~\cite{deepspeed3364}, developers' failure to provide library support for the M2 arm64 chip resulted in users being unable to install the library on devices with the M2 arm64 chip, causing program build failures. In ColossalAI Issue \textit{\#6058}~\cite{ColossalAI6058}, developers used modules specific to the Linux system in the code, which caused the program to fail during construction in the Windows environment due to the inability to import the modules.

\begin{figure}[h]
    \vspace{-11pt}
    \centering
    \includegraphics[width=0.9\columnwidth]{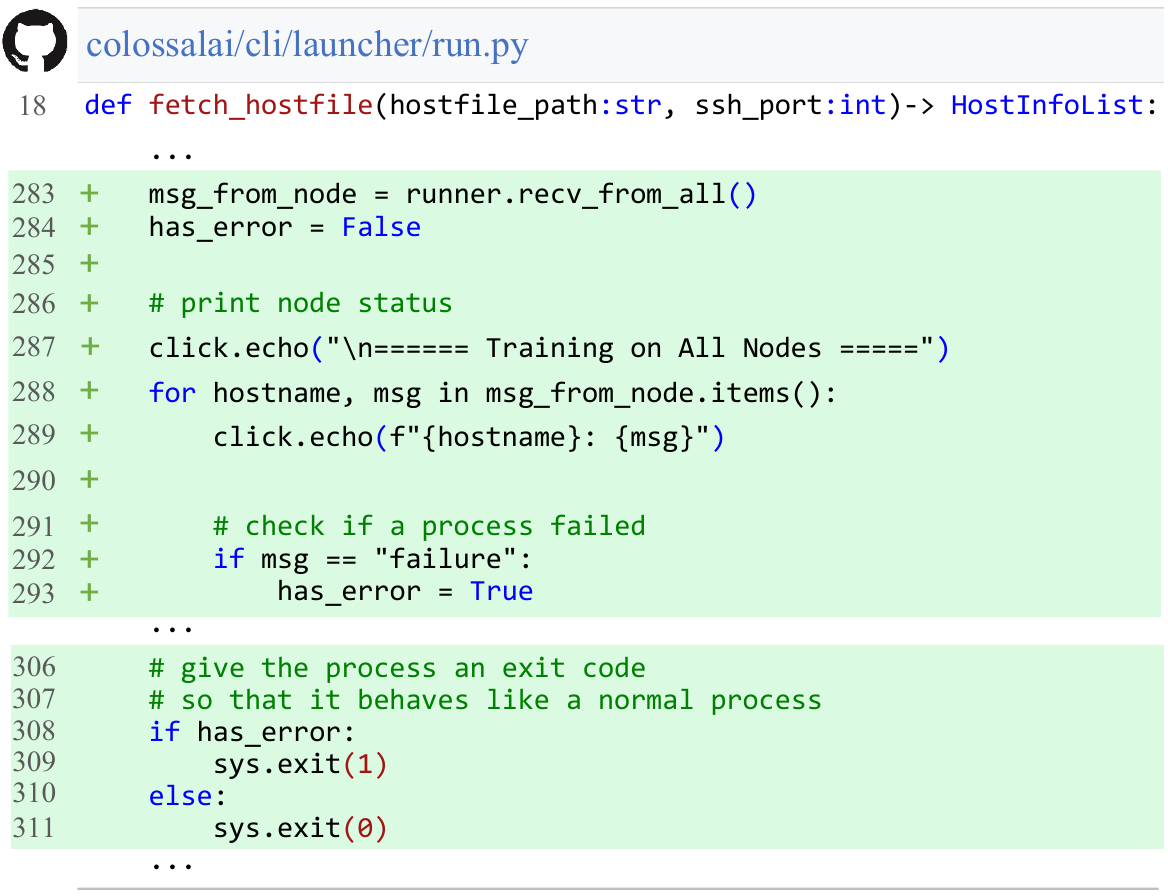} 
    \caption{An example of the distributed communication error.}
    \label{fig:bug_distributed_commmunication_error}
    \vspace{-11pt}
\end{figure}

\noindent\faBug\ \textbf{Distributed Communication Error:} This type of error is caused by improper handling of communication between distributed nodes by developers.
For example, in DeepSpeed Issue \textit{\#1995}~\cite{deepspeed1995}, developers mishandled termination signal logic across nodes, causing failed normal shutdown. In Colossal-AI Issue \textit{\#2447}~\cite{colossalai2447}, when a worker node encounters an error, it failed to transmit the error status to the master node through the communication mechanism. Therefore, the CLI (command-line interface) always displayed that the worker nodes were ``running normally.'' As shown in Figure \ref{fig:bug_distributed_commmunication_error}, the modified program checks for node errors via the code in lines 288-293: if any node returns ``\texttt{failure}'', it sets \texttt{has\_error} to \texttt{True}. If an error exists, the program returns 1, and the CLI will display a non-zero exit code (Lines 308-311).

\noindent\faBug\ \textbf{Incorrect Execution Order:} This type of error arises from improper sequencing of code blocks. For example, in DeepSpeed Issue \textit{\#6569}~\cite{deepspeed6569}, the update operation of the learning rate scheduler (\texttt{lr\_scheduler.step()}) was executed after the optimizer update (\texttt{optimizer.step()}), causing the optimizer to still use the default learning rate (e.g., 1e-3).

\begin{figure}[h]
    \centering
    \includegraphics[width=0.9\columnwidth]{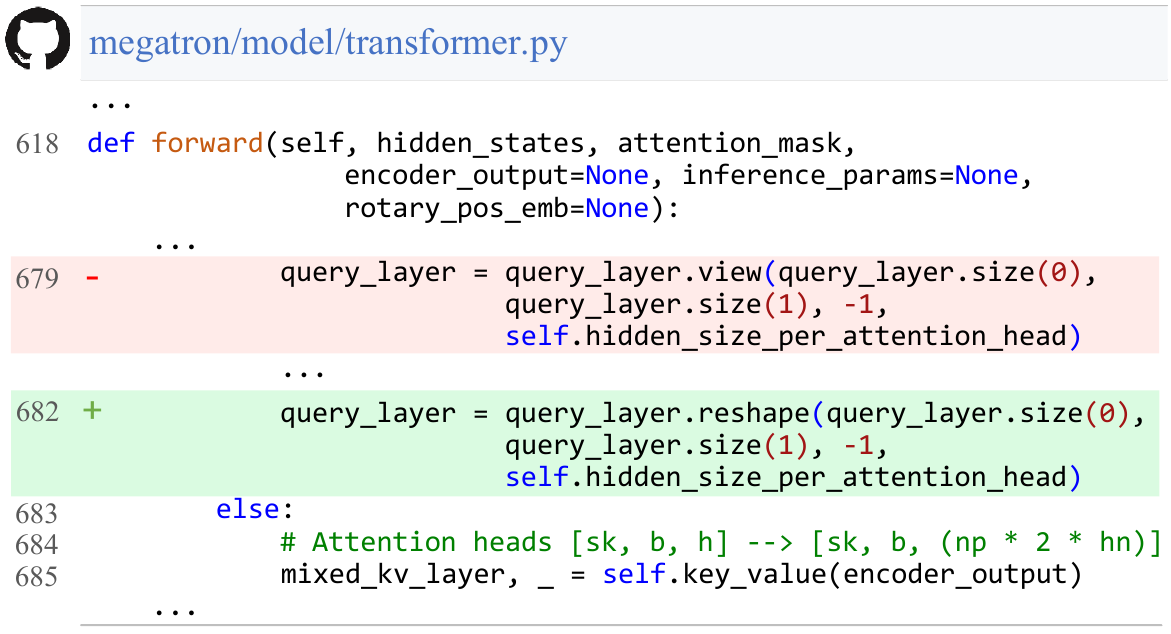} 
    \caption{An example of the distributed tensor error.}
    \label{fig:bug_distributed_tensor_error}
        \vspace{-11pt}
\end{figure}

\noindent\faBug\ \textbf{Distributed Tensor Error:} This error type arises from improper handling of tensors across distributed nodes, leading to computation failures or anomalies. In Megatron-LM Issue \textit{\#437}~\cite{megatronlm437}, as shown in Figure \ref{fig:bug_distributed_tensor_error}, when the user enables \texttt{pipeline-model-parallel-size=2} (pipeline parallelism with 2 device groups), the memory layout of tensors may become non-contiguous after cross-device transmission. The original code's \texttt{view()} operation (line 679) assumed contiguous memory, leading to failures in tensor processing across devices. The modified code uses the \texttt{reshape()} function (line 682), which can handle non-contiguous memory.

\begin{figure}[h]
    \vspace{-11pt}
    \centering
    \includegraphics[width=\columnwidth]{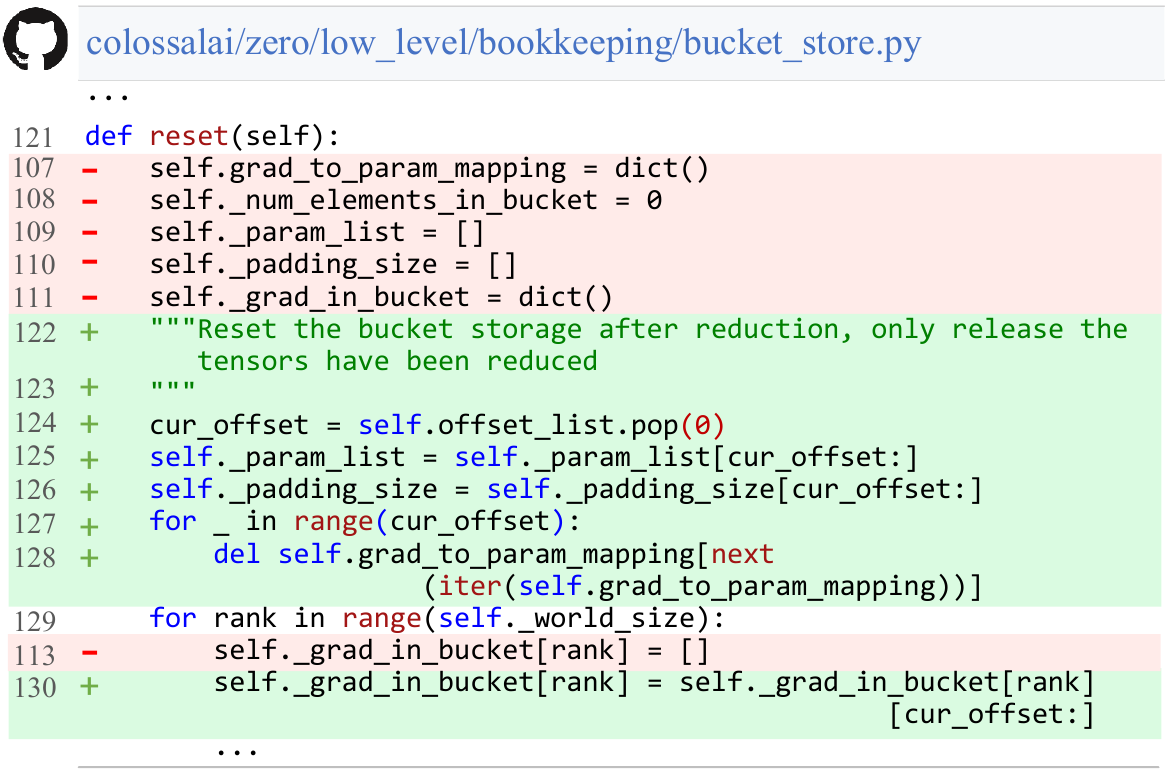} 
    \caption{An example of the concurrency error.}
    \label{fig:bug_concurrency_error}
    \vspace{-11pt}
\end{figure}

\noindent\faBug\ \textbf{Concurrency Error:} This type of error arises when developers fail to account for issues in concurrent execution scenarios. For example, in Colossal-AI Issue \textit{\#4402}~\cite{colossalai4402} (in Figure \ref{fig:bug_concurrency_error}) multiple processes  parallelize gradient computation, communication, and parameter updates. In the original code, once a process completes gradient communication, it calls reset to clear all parameters and gradients in the bucket (line 113). If other processes are still using these data, this leads to accessing invalid memory or empty lists, causing crashes or incorrect results. The modified code introduces a \texttt{cur\_offset} offset mechanism (lines 124-128): it only deletes parameters and gradients that have been used in communication while retaining unprocessed portions (line 130). 

\noindent\faBug\ \textbf{Incorrect Exception Handling:} This type of error arises from developers' flawed handling of program exceptions, primarily manifesting as inaccurate or missing error messages. In DeepSpeed Issue \textit{\#3669}~\cite{deepspeed3669}, the error message lacked the specific line number when an exception was thrown, complicating issue diagnosis. In ColossalAI Issue \textit{\#6019}~\cite{colossalai6019}, deviations in the calculation logic for missing keys during model state dictionary loading caused a discrepancy between error messages and actual key matching conditions, affecting the efficiency of issue localization.

As shown in Figure \ref{fig:root_cause_symptom_relationship}, almost all root causes are capable of triggering the four most prevalent symptoms—crash, incorrect functionality, build failure, and poor performance. This suggests that identifying the underlying root cause proves particularly challenging when these symptoms occur, as locating the root cause becomes challenging due to the wide range of potential root causes associated with each symptom. 
In addition, CI/CD failure is primarily linked to API misuse, configuration error, API incompatibility, and incorrect condition, whereas hang predominantly stems from environmental incompatibility, distributed communication errors, and concurrency error.

\begin{center}
\vspace{-15pt}
    \resizebox{\linewidth}{!}{
\begin{tabular}{l!{\vrule width 1pt}p{0.9\columnwidth}}
    \makecell{{\LARGE \faLightbulbO}}  &\textbf{Finding 2.} API misuse, incorrect implementation, configuration error, and precondition missing emerge as the four most prevalent root causes. The distributed nature of these frameworks gives rise to four distinctive categories of root causes: allocation strategy error, distributed communication error, distributed tensor error, and concurrency error. Almost all root causes can lead to the four most prevalent symptoms—crash, incorrect functionality, build failure, and poor performance. 
\end{tabular}}
\vspace{-8pt}
\end{center} 

\subsection{RQ3: Challenges in Root Cause Identification and Bug Fixing}

\begin{figure}[t]
    \centering
    \includegraphics[width=\columnwidth]{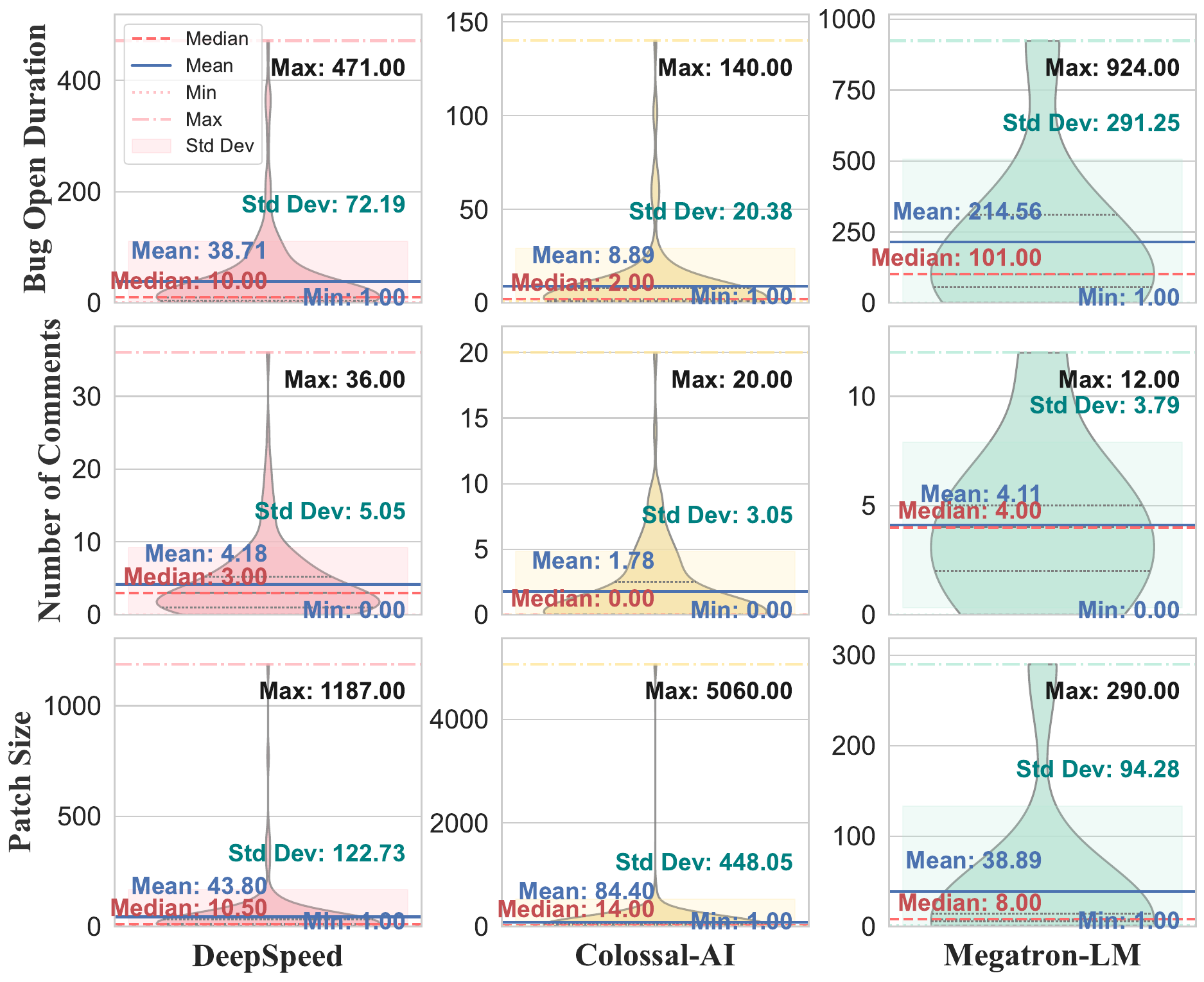} 
    \caption{The violin picture of bug open duration, number of comments, and patch size}
    \label{fig:violin_picture}
\vspace{-15pt}
\end{figure}

Figure \ref{fig:violin_picture} presents the violin plots of bug open duration, number of comments, and patch size. Colossal-AI exhibits shorter bug open durations than DeepSpeed and Megatron-LM, possibly because developers promptly address issues upon reporting, which correlates with fewer comments per issue. However, all three frameworks show similar patch sizes, with medians ranging from 8 to 14 LOC. Following previous work~\cite{wang2021exploratory,yang2025towards}, we analyze bug-labeled issues that remain open for over one month and receive more than ten comments, comprising approximately 6\% of all cases. These prolonged issues reflect substantial debugging effort. Through analyzing these cases, we identify three major challenges in pinpointing the root causes of bugs in the frameworks. 

\textbf{The Disconnect Between Bug Symptoms and Causes}: The difficulty of locating bugs often increases significantly when symptoms show no direct correlation with root causes. For example, in DeepSpeed Issue \textit{\#2911}~\cite{deepspeed2911}, users reported bf16/fp32 mode system logs showed loss scaler warnings despite the feature being off. Initially, the issue proposer suspected a warning module anomaly. However, joint analysis by community developers revealed the core problem: fp16 data was transmitted in a bf16 state during distributed communication, causing infinite gradient values that triggered the warning.

%
\textbf{High Resource Overhead for Bug Reproduction}: 
Reproducing bugs in LLM training with distributed training and inference frameworks often demands substantial computational resources. For instance, in DeepSpeed Issue \#\textit{3481}~\cite{deepspeed3481}, an error is triggered exclusively during the training of a large-parameter model on 8 A100 GPUs (80GB) and 1024GB RAM when initializing the optimizer states and attempting to pin gradient partitions to memory (\texttt{pin\_memory}). Subsequently, another user attempted to replicate the bug using a setup with limited hardware resources but found the model became stuck for an hour during training, failing to reproduce the error.
During the discussion, a DeepSpeed developer acknowledged that memory pinning operations for only large-parameter models impose extremely high memory demands. In other words, insufficient device memory prevents successful completion of these operations, thereby hindering bug reproduction.

\textbf{Low-Level Bugs in the Distributed Training and Inference Frameworks}: This challenge refers to low-level bugs rooted in the fundamental architectural design of distributed training and inference frameworks. 
For example, in DeepSpeed Issue \textit{\#2449}~\cite{deepspeed2449}, DeepSpeed Stage 3 failed to load checkpoints when no optimizer was configured during inference. From the perspective of conventional inference workflows, the optimizer is not a necessary configuration item, prompting developers to remark with surprise, \textit{“This is extremely counter-intuitive.”} Through subsequent in-depth and sustained discussions, developers gradually identified the root cause of this bug as a flaw in the library's architectural design. Due to the unreasonable strong coupling between the inference engine and the optimizer, even in inference scenarios where optimization operations are unnecessary, the optimizer must still be configured. This architectural irrationality significantly complicated the root cause identification process.


To understand the reasons behind the substantial fixing efforts required for bugs in distributed training and inference frameworks, we follow previous work~\cite{wang2021exploratory,yang2025towards} and manually review PRs containing the top 20\% of patches, each exceeding 50 LOC in size. We identify three key challenges in bug fixing.


\textbf{Fixing Low-Level Bugs in the Frameworks}:
Modifying low-level operations within a distributed training and inference framework and their associated functionalities presents significant challenges. For example, in Colossal-AI Issue \textit{\#3471}~\cite{colossalai3471}, user reports showed \texttt{ColoTensor} lacked in-place operation support. As a tensor type in Colossal-AI for memory optimization in distributed training, \texttt{ColoTensor} serves as a fundamental data type in the framework. To enable in-place operations, developers modified its underlying implementation, refactored all library components using \texttt{ColoTensor}. This effort spanned nearly five months, involved modifications to 37 production code files, and edits to 1,775 LOC.


\textbf{Fixing Algorithm Implementation Logic in Distributed Training and Inference}: Fixing functional bugs or algorithmic logic errors in a library often requires coordinated modifications to related modules and test cases, significantly increasing development complexity and demanding substantial time for logical reorganization and task coordination. For example, in
DeepSpeed Issue \textit{\#1950}~\cite{deepspeed1950}, user reported inconsistencies in positional encoding and fusion pattern handling within the distributed attention mechanism caused non-reproducible inference discrepancies. Developers swiftly patched the Transformer module's attention implementation (e.g., positional encoding, matrix operation, attention flow), overhauling both C++ logic and Python interfaces. After approximately two weeks of work and 1155 LOC changes, the issue was resolved.


\textbf{Fixing Bugs across Multiple Components}: Fixing bugs spanning multiple modules poses significant challenges. Developers must ensure functionality integrity while preventing collateral impacts, requiring deep module knowledge and much collaborative effort. In Colossal-AI Issue \textit{\#5187}~\cite{colossalai5187}, A custom shard policy caused pipeline layer distribution inconsistencies. Resolving this involved coordinating across model sharding, hybrid parallelism, pipeline engine, and sharding strategy modules. Five developers spent approximately three and a half months and modified 14 production code files to complete the fix.

\begin{center}
\vspace{-15pt}
    \resizebox{\linewidth}{!}{
\begin{tabular}{l!{\vrule width 1pt}p{0.9\columnwidth}}
    \makecell{{\LARGE \faLightbulbO}}  &\textbf{Finding 3.} The challenges of root cause identification and bug fixing in distributed training and inference frameworks mainly stem from the disconnect between bug symptoms and causes, high resource overhead for reproduction, low-level framework bugs and their fixes, algorithm implementation logic fixes, and cross-component bug fixes. 
\end{tabular}}
\vspace{-8pt}
\end{center} 

\subsection{RQ4: Common Fixing Strategy}


We discover that 48\% of bug fixes involve no more than 10 LOC, suggesting simple modifications suffice. Through manual inspection of these patches, we identify several small-LOC fixing strategies that could potentially be automated.

\noindent\faWrench\ \textbf{Conditional Logic Optimization}: This strategy enhances programs’ capability to handle potential scenarios by optimizing conditional logic. For example, in DeepSpeed \textit{\#3777}~\cite{deepspeed3777}, developers introduced a check for the \texttt{requires\_grad} attribute in model parameters to avoid unnecessary gradient memory allocation.

\noindent\faWrench\ \textbf{Parameter Handling Enhancement}: This type of strategy introduces additional parameters to the program or enhances parameter handling, providing developers with more comprehensive configuration options to adapt to actual requirements flexibly and effectively resolve issues. For example, in Colossal-AI Issue \textit{\#4903}~\cite{colossalai4903}, developers addressed a library update conflict by adding the \texttt{**kwargs} parameter to the \texttt{attention\_forward} function. This allowed the function to retain its original parameter structure while dynamically accepting undeclared keyword arguments, thus enhancing compatibility.

\noindent\faWrench\ \textbf{API Replacement}:
This strategy addresses issues arising from context-specific API limitations by replacing employed APIs with more suitable alternatives that maintain functionality. For example, in Megatron-LM \textit{\#437}~\cite{megatronlm437}, developers replaced the \texttt{view} API with \texttt{reshape} for tensor operations, 
a more robust API capable of handling non-contiguous tensors generated during \textit{Group Query Attention} operations, thereby ensuring tensor correctness in parallel computation.

\noindent\faWrench\ \textbf{Configuration Adjustment}: This type of strategy corrects functional anomalies caused by configuration deviations by adjusting the program's configuration parameters. For example, in Colossal-AI Issue \textit{\#4027}~\cite{colossalai4027}, the code defaultly configured the communication bucket size between distributed nodes as 12T, which significantly impacted runtime efficiency. Developers later adjusted it to 12MB with dynamic optimization based on runtime states. 

\noindent\faWrench\ \textbf{Version Compatibility Handling}: This strategy proactively mitigates issues stemming from version disparities by incorporating checks for third-party library versions or hardware configurations. A notable example is found in Colossal-AI Issue \textit{\#4829}~\cite{colossalai4829}, where developers addressed PyTorch version incompatibilities by implementing a conditional version check. Specifically, they fixed the code to dynamically import compatible scheduler base classes: \texttt{\_LRScheduler} for PyTorch versions below 2.0 and \texttt{LRScheduler} for 2.0 and above, depending on the detected runtime environment. 

\begin{center}
\vspace{-15pt}
    \resizebox{\linewidth}{!}{
\begin{tabular}{l!{\vrule width 1pt}p{0.9\columnwidth}}
    \makecell{{\LARGE \faLightbulbO}}  &\textbf{Finding 4.} Nearly half (48\%) of bug fixes involve minimal code changes ($\leq$ 10 LOC) and can be addressed through five simple strategies: conditional logic optimization, parameter handling enhancement, API replacement, configuration adjustment, and version compatibility handling.
\end{tabular}}
\vspace{-8pt}
\end{center} 

\section{Discussion}
\label{sec: discussion}
\subsection{Implications}

We give some practical implications for three primary stakeholders in the distributed training and inference ecosystem: 

\noindent \textbf{Implications for Framework Developers:} (1) A large portion of bugs stem from API misuse, incorrect implementations, configuration error, and precondition missing. To reduce such issues, developers should implement rigorous input validation and incorporate defensive checks at API boundaries to catch incorrect usage early. (2) Many bugs ~\cite{deepspeed5603, deepspeed5534, deepspeed4853, colossalai4829, colossalai2938} are triggered by incompatibilities with specific versions of PyTorch or CUDA. Maintaining version-aware test matrices within CI/CD pipelines and explicitly documenting compatibility requirements can help avoid the problems. (3) From a debugging perspective, reproducing certain bugs often requires access to large-scale GPU clusters, creating substantial barriers to timely diagnosis and resolution. To mitigate this, we recommend supporting mock modes or lightweight execution paths (e.g., simulating multi-node setups on a single machine) to facilitate local reproduction and faster debugging. (4) The disconnect between observable symptoms and their root causes and low-level or cross-component bugs make bug localization particularly challenging. Providing structured, verbose, and consistent runtime logs, including node identifiers, tensor shapes, and communication statuses, can significantly enhance diagnosability and reduce the debugging burden on developers.

\noindent \textbf{Implications for Framework Users:} 
By analyzing GitHub repositories through dependency files (e.g., \texttt{requirements.txt}), we identify widespread adoption of distributed training and inference frameworks across the open-source ecosystem. For example, DeepSpeed alone is used by 375 projects with over 100 GitHub stars, including 147 projects with over 1,000 stars and 22 exceeding 10,000 stars. While Colossal-AI and Megatron-LM show comparatively lower adoption, they are still used in 15 and 10 popular projects (with over 100 stars), respectively. Importantly, the majority of these projects focus on LLM development. These statistics underscore the critical impact that bugs in distributed training and inference frameworks can have on downstream applications. However, due to the modular and layered architecture of these frameworks, their complex and opaque call networks often hinder users from identifying which APIs or components are affected by known bugs. Our study further reveals that some bugs only manifest under specific runtime conditions. For instance, DeepSpeed Issue \#\textit{3481}~\cite{deepspeed3481} only triggers when training large-parameter models, making it difficult for developers to anticipate failures through static inspection alone. Such context-dependent behavior limits the effectiveness of conventional code review and static analysis techniques in identifying all buggy paths. Therefore, we recommend that framework users—especially those building large-scale LLM systems—adopt advanced analysis tools that combine static code structure with dynamic execution traces. Constructing hybrid call graphs that integrate static call relationships with runtime behaviors can help users map their application code to potentially buggy framework components. Furthermore, incorporating runtime context (e.g., input data size, model parameter count) into these analyses can improve precision in identifying when and where a bug is likely to occur. Such strategies can help users proactively assess risk and mitigate failures before deployment.

\noindent \textbf{Implications for Framework Researchers:} 
While LLM-based methods have demonstrated strong performance in program repair tasks~\cite{hossain2024deep} (e.g., SWE-bench benchmark~\cite{jimenez2023swe}), their applicability to distributed training and inference frameworks remains largely underexplored. (1) Nearly half of the framework bugs being resolved with minimal code changes suggests strong potential for repair approaches that combine fix templates or patterns with LLMs. In this paradigm, templates mined from historical patches provide structural guidance, while LLMs generate context-aware code to complete the fix. (2) The unique failure modes in these frameworks, such as allocation strategy error, distributed communication error, and distributed tensor error, demand specialized diagnostic and reasoning capabilities that go beyond conventional repair techniques. Integrating LLMs with RAG could help bridge this gap. For example, when encountering an allocation strategy error, a RAG-enhanced approach could retrieve relevant design documentation or previously reported issues involving memory sharding, thereby enabling more precise diagnosis and repair suggestions. (3) Our study further highlights that complex bugs---particularly those involving non-obvious symptom-cause relationships, flawed algorithmic logic, cross-component interactions, or low-level data structures such as \texttt{ColoTensor}~\cite{colossalai3471}---remain difficult to diagnose and resolve. Addressing such bugs requires deep architectural understanding and multi-hop reasoning. Recent advances in reasoning-oriented LLMs (e.g., GPT-o1~\cite{OpenAIO1}, DeepSeek-R1~\cite{guo2025deepseek}) offer promising avenues for simulating the deliberative processes of experienced developers. We envision future research exploring the integration of these models with program analysis tools to create hybrid approaches capable of addressing semantically complex bugs in distributed frameworks.

\subsection{Threats of Validity}

Our findings may have limited generalizability to all distributed training and inference frameworks due to two factors. First, we focus exclusively on open-source frameworks, whereas industrial-scale LLM training often relies on proprietary, closed-source frameworks (e.g., DeepSeek’s in-house distributed training and inference system~\cite{liu2024deepseek}). The inaccessibility of these closed frameworks introduces bias, as their bug characteristics (e.g., root causes, fix strategies) may differ from those of open-source counterparts. Second, the number of analyzed bugs is comparable to or smaller than that in recent studies on traditional machine learning libraries (e.g., Yang \textit{et al}.~\cite{yang2025towards} analyze 202 performance bugs from seven popular DS libraries) and deep learning libraries (Chen \textit{et al}.~\cite{chen2023toward} analyze 250 bugs per deep learning library), primarily because LLM distributed training and inference frameworks are nascent and bug accumulation and documentation evolve slowly. To mitigate these threats, we select widely adopted open-source frameworks with active developer communities, ensuring our sample reflects the most representative open-source ecosystem. Future work will expand data collection to include emerging popular distributed training and inference frameworks and extend the time span to validate the generalizability of our findings and investigate bug evolutionary patterns. Another validity threat stems from the manual annotation process, which introduces risks of subjectivity. To address this, two authors independently label the data using an open-coding scheme consistent with prior empirical studies. Any labeling discrepancies are resolved through iterative discussions with an experienced independent third
researcher, ensuring inter-rater reliability and consistent interpretation. In addition, since we cannot access direct measures of bug localization and fixing difficulty (e.g., developer time spent debugging), we employ surrogate metrics consistent with prior research~\cite{zaman2012qualitative, liu2014characterizing, yang2025towards}. Specifically, bug open duration and issue comment count serve as proxies for root cause identification complexity, while patch size quantifies fixing difficulty.

\section{Related Works}

\label{sec:related work}

\noindent \textbf{Traditional Parallel/Distributed Training Bugs.} 
Empirical studies on bugs in parallel/distributed systems or programs have primarily focused on traditional shared-memory architectures and general-purpose distributed systems~\cite{abbaspour2018concurrency, gao2018empirical, li2013characteristic}. However, distributed training and inference frameworks for LLMs differ significantly from these paradigms in both hardware architecture and program characteristics. Unlike shared-memory parallel systems where processors communicate via direct memory access~\cite{gelado2010asymmetric}, LLM distributed training and inference relies on device-to-device communication protocols across GPUs with isolated memory spaces.
These unique characteristics fundamentally differentiate LLM distributed training and inference bugs from those in traditional parallel/distributed training systems or programs. Therefore, we do not elaborate on existing bug analysis studies of traditional paradigms. To date, only Liu \textit{et al}.~\cite{liu2023rise} analyzed 1,131 real-world developer issues in distributed training reported on Stack Overflow and GitHub, focusing on challenges encountered when using TensorFlow, PyTorch, Keras, and Horovod to develop deep learning programs for distributed training. In contrast, our study explicitly targets inherent bugs within the distributed training and inference frameworks themselves. Moreover, the frameworks we analyze are those more commonly adopted for LLM distributed training and inference.

\noindent \textbf{Machine/Deep Learning Bugs.} 
Existing studies on machine and deep learning bugs broadly categorize bugs into two classes: program bugs and framework bugs. Research on program bugs~\cite{zhang2018empirical,islam2019comprehensive,humbatova2020taxonomy,cao2022understanding,wang2023compatibility,hong2024investigating,morovati2024bug} primarily focuses on issues arising from the incorrect use of machine/deep learning framework APIs. These bugs stem from user errors rather than bugs in the underlying implementation of the frameworks. For example, Islam \textit{et al}.~\cite{islam2019comprehensive} investigated bugs in deep learning programs implemented using Caffe, Keras, TensorFlow, Theano, and Torch, analyzing bug types, root causes, impacts, and the stages in the deep learning pipeline that are most prone to bugs. In contrast, several recent studies~\cite{jia2020empirical, jia2021symptoms, yang2022comprehensive, long2022reporting, ho2023empirical, li2023understanding, tambon2024silent, chen2023toward, lai2024security, yang2025towards} have examined bugs with the machine/deep learning frameworks themselves, exploring symptoms, root causes, and bug fixing patterns for mainstream frameworks. For instance, Chen \textit{et al}.~\cite{chen2023toward} conducted the most extensive analysis to date, examining 1,000 bugs inside TensorFlow, PyTorch, MXNet, and DL4J, investigating the root causes and symptoms of these bugs, as well as identifying which levels in deep learning frameworks are more susceptible to bugs. Our study reveals some overlap with the findings of Chen \textit{et al}.\cite{chen2023toward}: incorrect implementation, configuration error, and API misuse are common root causes in both deep learning frameworks and distributed training and inference frameworks. However, we uniquely identify distributed-specific root causes, such as allocation strategy errors and distributed communication errors. Additionally, while Chen \textit{et al}.\cite{chen2023toward} reports crash and hang as dominant symptoms, we find that hang is rare in distributed training and inference frameworks. Importantly, Chen \textit{et al}.~\cite{chen2023toward} does not analyze the challenges of root cause identification and bug fixing, nor does it discuss common low-effort fixing strategies—both of which are key contributions of our study.

\section{Conclusion}
\label{sec: conclusion}

Our empirical study of 308 bugs across DeepSpeed, Megatron-LM, and Colossal-AI reveals that most bugs manifest as crash, incorrect functionality, or build failure, frequently caused by API misuse, incorrect implementation, and configuration error. In addition, the distributed design introduces unique bug root causes, such as allocation strategy error and distributed communication error. We identify several key challenges in debugging complex bugs, including the disconnect between observed symptoms and root causes, the high cost of reproducing bugs, and the involvement of low-level or cross-component bugs. Despite this complexity, 48\% of bugs can be resolved with minimal code changes, often by applying simple fixes such as conditional logic optimization or configuration adjustments. Finally, based on our findings, we provide a set of actionable guidelines to improve the debugging and repair practices for distributed training and inference frameworks and their dependent LLM projects. 

\bibliographystyle{IEEEtran}
\bibliography{main.bib}

\end{document}